\documentclass[sigconf]{acmart}

\usepackage{booktabs} 

\usepackage[utf8]{inputenc}

\usepackage{enumerate}

\usepackage{amssymb}
\usepackage{makecell}
\usepackage{graphicx}

\setcopyright{rightsretained}

\acmDOI{10.475/123_4}

\acmISBN{123-4567-24-567/08/06}

\acmConference[RecSys'18]{ACM Conference on Recommender Systems}{October 2018}{Vancouver, Canada}
\acmYear{2018}
\copyrightyear{2018}

\acmArticle{4}
\acmPrice{15.00}

\begin{document}
\title{CF4CF: Recommending Collaborative Filtering algorithms using Collaborative Filtering}

\author{Tiago Cunha}
\orcid{0000-0003-0564-6129}
\affiliation{%
  \institution{Faculdade de Engenharia da Universidade do Porto}
  \streetaddress{Rua Dr. Roberto Frias}
  \city{Porto} 
  \country{Portugal} 
  \postcode{4200-465}
}
\email{tiagodscunha@fe.up.pt}

\author{Carlos Soares}
\affiliation{%
  \institution{Faculdade de Engenharia da Universidade do Porto}
  \streetaddress{Rua Dr. Roberto Frias}
  \city{Porto} 
  \country{Portugal} 
  \postcode{4200-465}
}
\email{csoares@fe.up.pt}

\author{André C.P.L.F. de Carvalho}
\affiliation{%
  \institution{Universidade de São Paulo, ICMC }
  \streetaddress{Rua Trabalhador Sancarlense}
  \city{São Carlos}
  \state{São Paulo}
  \country{Brasil}}
\email{andre@icmc.usp.br}

\renewcommand{\shortauthors}{Cunha et al.}

\begin{abstract}

Automatic solutions which enable the selection of the best algorithms for a new problem are commonly found in the literature. One research area which has recently received considerable efforts is Collaborative Filtering. Existing work includes several approaches using Metalearning, which relate the characteristics of datasets with the performance of the algorithms. This work explores an alternative approach to tackle this problem. Since, in essence, both are recommendation problems, this work uses Collaborative Filtering algorithms to select Collaborative Filtering algorithms. Our approach integrates subsampling landmarkers, which are a data characterization approach commonly used in Metalearning, with a standard Collaborative Filtering method. The experimental results show that CF4CF competes with standard Metalearning strategies in the problem of Collaborative Filtering algorithm selection.

\end{abstract}

%
%



\maketitle

\section{Introduction}

The algorithm selection problem for Collaborative Filtering (CF)~\cite{Shi2014} has been investigated so far via Metalearning (MtL)~\cite{Adomavicius2012,Ekstrand2012,Griffith2012,Matuszyk2014,Cunha2016,Cunha2017, Cunha:2017:MCF:3109859.3109899}. The problem is modeled using a set of features (i.e., metafeatures) to describe the problem domain and the performance of algorithms according to a specific measure to describe the behavior of algorithms. Afterwards, learning algorithms are used to learn the mapping between the metafeatures and the performance, effectively achieving a model (i.e. metamodel) which can be used to predict the best algorithms for a new problem. 

However, the definition of suitable metafeatures is a hard problem. This is specially difficult in the CF problem, where there is no clear separation between independent and dependent variables. So far, there have been several examples of statistical and/or information-theoretical approaches~\cite{Adomavicius2012,Ekstrand2012,Griffith2012,Matuszyk2014,Cunha2016} and even landmarking approaches~\cite{Cunha2017}, which have produced interesting results. However, the merits of metafeatures continue to be questioned, since it is difficult to understand whether they actually contain useful informative or whether the results are dictated by noise or chance. Hence, we look towards another approach, which does not use metafeatures explicitly to train the metamodel. 

The approach proposed in this work is to use CF algorithms to select CF algorithms, which we name CF4CF. The problem is addressed by considering users and items as the datasets and algorithm, respectively. The performance of all algorithms on a particular dataset are leveraged and converted into ratings. Thus, a proper rating matrix can be built using performance data only. Then a CF algorithm can be used to create a metamodel, which will allow to predict the best ranking of algorithms for a new problem. Specifically in the prediction step, when no data is available regarding the algorithm performance, CF4CF uses subsampling landmarkers (performance estimations on a sample of the original dataset) to obtain initial ratings. CF4CF is then responsible to predict the remaining ratings and convert the outcome into a ranking of algorithms. 

As far as the authors know, this paper's contribution - CF4CF - is the first approach to use CF algorithms to recommend CF algorithms. Furthermore, this is also the first attempt of CF algorithm selection which does not explicitly use metafeatures in the trained model. Beyond the interestingness of proving the ability to tackle the algorithm selection problem without metafeatures, this work is particularly important because it allows to compare the merits of traditional MtL and the novel CF4CF approaches. To this end, this work compares the merits of metalevel accuracy and impact on the baselevel for both learning strategies and shows that CF4CF is a suitable alternative for algorithm selection, having proved to be perform equally or better than traditional MtL.

This document is organized as follows: Section~\ref{sec:meta_cf} presents the related work on Metalearning for CF; Section~\ref{sec:process} presents the core contributions of this work: CF4CF and the unified evaluation framework, while Section~\ref{sec:experimental_setup} explains the experimental procedure. In Section~\ref{sec:results}, the proposed approach is evaluated and discussed and Section~\ref{sec:conclusions} presents the conclusions and future work tasks.

\section{Related Work}\label{sec:meta_cf}

Although the use of MtL for CF has already been investigated~\cite{Adomavicius2012,Ekstrand2012,Griffith2012,Matuszyk2014},
the approaches proposed have limited scope: the set of datasets, recommendation algorithms and metafeatures studied is always suitable, but never complete. An extensive overview of their positive and negative aspects can be seen in a recent survey~\cite{Cunha2018128}. More recent work in CF algorithm selection has extended the contributions to the area, in particular with regards to the metafeatures considered, which systematize the data characteristics used in earlier works~\cite{Cunha2016}. This work, which we consider as the state of the art in CF algorithm selection, proposes a systematic approach for metafeature extraction. It leverages a framework which requires three main elements: object $o$, function $f$ and post-function $pf$. The framework applies a function to an object and, afterwards, the post-function to the outcome in order to derive the final metafeature. Thus, any metafeature can be represented as: $\{o.f.pf\}$~\cite{Pinto2016}.

The objects to be used in the framework are CF's rating matrix $R$, and its rows $U$ and columns $I$. The functions $f$ considered to characterize these objects are: original ratings (\textit{ratings}), count the number of elements (\textit{count}), mean value (\textit{mean}) and sum of values (\textit{sum}). The post-functions $pf$ are maximum, minimum, mean, standard deviation, median, mode, entropy, Gini index, skewness and kurtosis. Additionally, it includes the number of users, items, ratings and the matrix sparsity. This results in 74 metafeatures which were reduced by correlation feature selection, ending up with: \textit{nusers}, \textit{R.ratings.kurtosis}, \textit{R.ratings.sd}, \textit{I.count.kurtosis}, \textit{I.count.min}, \textit{I.mean.entropy},  \textit{I.sum.skewness}, \textit{U.sum.entropy}, \textit{U.mean.min},  \textit{sparsity}, \textit{U.sum.kurtosis}, \textit{U.mean.skewness}. As an example, \textit{R.ratings.kurtosis} represents the kurtosis of the distribution of all ratings in matrix $R$.

\section{CF4CF}\label{sec:process}

\begin{figure}[!ht]
  \centering
    \includegraphics[width=.9\linewidth]{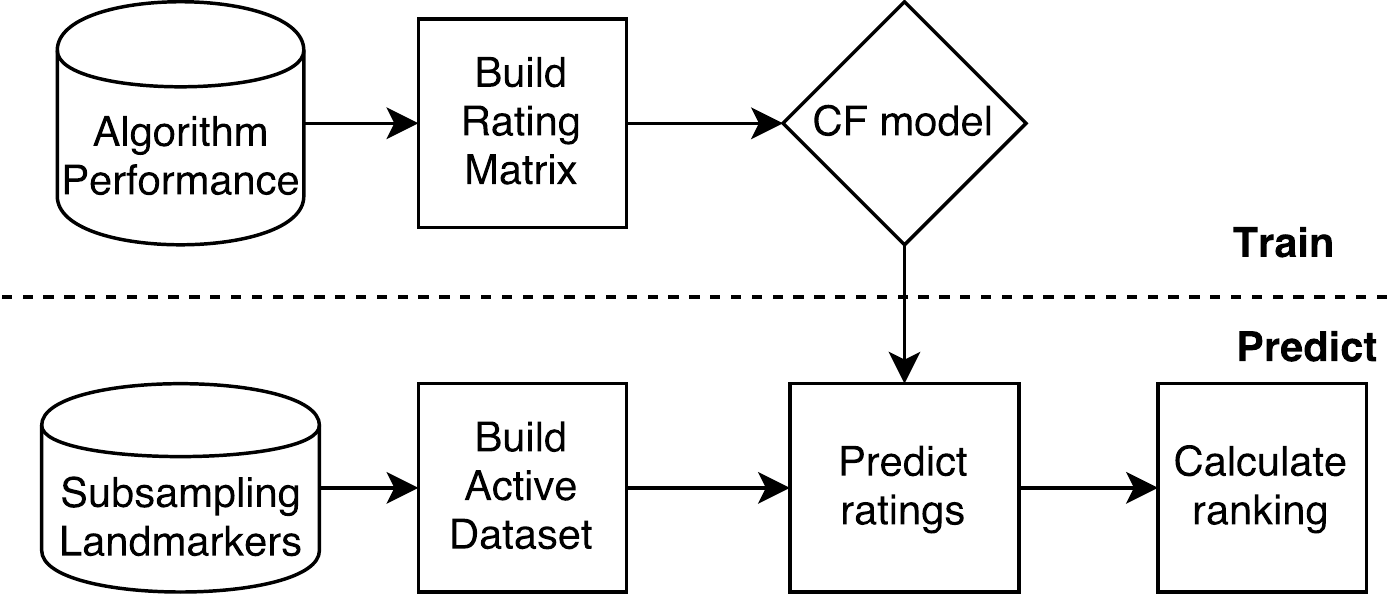}
      \caption{Overview of the CF4CF procedure.}
      \label{fig:cf4cf}
\end{figure}

This paper introduces a novel approach to tackle the CF algorithm selection problem, named CF4CF. Figure~\ref{fig:cf4cf} presents the procedure. 

Notice the process is organized in two main steps: train and predict. The training stage leverages the algorithm performance data, builds a rating matrix and trains a CF model. In the prediction stage, algorithm performance from subsampling landmarkers is transformed leveraged to create the initial ratings of the active dataset. The active dataset is then submitted to the previously trained CF model to obtain ratings for the missing algorithms. Afterwards, the final ranking of algorithms is calculated. The next sections will present in detail the steps exposed in the previous overview.

\subsection{Build the Rating Matrix}

Recall that CF requires three elements: users, items and ratings. As this work aims at recommending CF algorithms for CF datasets, the natural adaptation is to consider the users and items as datasets and algorithms, respectively. Hence, to build the rating matrix $R ^{D \times A}$ we consider the set of datasets $D$ where each dataset $ d_i \in D$ and the set of algorithms $A$ where each algorithm $ a_j \in A$. To complete the matrix, one needs to provide the ratings available. However, in the algorithm selection problem there is not an explicit assignment of ratings by each dataset to the algorithms. To solve this issue, we model the preferences using the performance of algorithms on the datasets. The idea is to leverage how good the algorithm is for a particular dataset as the preference it holds for the same dataset.

Our approach works by converting the rankings into ratings. This conversion allows to take advantage of CF algorithms in a straightforward way. Formally, consider a ranking of algorithms $R_{d_i} = (a_j)_{j=1}^{M}$ for a specific dataset $d_i$. Such ranking is created by sorting the algorithms in decreasing order of performance. To convert the ranking $R_{d_i}$ into a specific ratings scale $S \in [s_{min},s_{max}]$, the following transformation $f$ is applied to each position $j$:
\begin{equation}
f (R_{d_i},j) = \frac{(S_{max} - S_{min}) (M - j)}{M - 1} + S_{min} 
\label{eq:conversion}
\end{equation}
The rating values are then $R_{d_i,a_j} = f(R_{d_i},j)$. The matrix is completed by converting all rankings of algorithms for all datasets. 

\subsection{Train the CF model}

Notice the previous step outputs a complete rating matrix, since we have a preference for all datasets towards all algorithms. Although CF4CF uses a complete matrix, which is not the case in most CF problems, all CF algorithms available can be used in CF4CF. In the works case scenario, one just needs to sample the rating matrix to create missing data for algorithms such as Matrix Factorization to be able to operate. This is in fact a major advantage: since CF does not require all ratings to be provided, then it is theoretically possible to achieve good performance with less information than what is required by MtL, which may translate into significant saves in computational resources. 
The experimental procedure will assess these assumptions by varying the parameter $N_{ratings}$, which refers to the number of ratings sampled by dataset to build the matrix. 

\subsection{Build the Active Dataset}

Having the model built, one moves now to the prediction stage. However, due to domain constraints, one must introduce changes to the traditional prediction procedure. Recall that if a new dataset is considered, it is reasonable to assume that there is no performance estimate for any algorithm. In this case, CF4CF cannot properly work since it would have no data to provide the CF model. This work proposes to deal with this problem using subsampling landmarkers, which consists in estimating the algorithm performance on a small data samples and use them as initial input for the CF model.

Thus, in order to build the active dataset representation, this procedure leverages the subsampling landmarkers and processes them via sampling and rating conversion procedures. Formally, let us consider the complete ranking of algorithms $SL_{d_i} = (a_j)_{j=1}^{M}$ for a specific dataset $d_i$, obtained from subsampling landmarkers rather than the original performance values. Since we aim to use some of these values to serve as initial ratings for the CF model, we first sample the ranking $SL_{d_i}$. Considering how the number of ratings provided directly affects the performance of CF models, it is important to understand the effect of sampling different amounts of ratings. We address this issue by  using a parameter $N_{SL} \in [1,...M-1]$ in our experiments. Lastly, the sampled ranking is converted into ratings, also using Equation~\ref{eq:conversion}.

\subsection{Predict Ratings and Calculate Ranking}

Having obtained the active dataset representation $SL_{d_i}$, one uses the previously trained CF model to obtain the predictions for the remaining algorithms, represented as $\hat{R}_{d_i}$. Notice that CF algorithms only considers items for which the active user has not provided any feedback towards. Hence, in our case, CF will produce ratings for the remaining algorithms in a straightforward way. 

Notice however the algorithm selection problem requires a complete ranking of algorithms to be predicted. To tackle this issue, we propose to aggregate the predictions with the initial ratings. Hence, the full ratings predicted are provided by $R_{d_i} = <\hat{R}_{d_i}, SL_{d_i}>$. 

At this point, the only step remaining is to convert the ratings into rankings. To do so,  one sorts the ratings in decreasing order of importance and replaces them by the respective ranking position. By fixing the algorithm positions, one ensures a representation which allows to directly use ranking accuracy measures and, by extension, to compare CF4CF with MtL.

\section{Experimental setup}\label{sec:experimental_setup}

\subsection{Baselevel}

The baselevel component is concerned with the traditional CF problem and it is exactly the same for both CF4CF and MtL. Here, several dimensions are considered: datasets, algorithms and evaluation measures. The 38 datasets used come from different domains, namely Amazon Reviews, BookCrossing, Flixter, Jester, MovieLens, MovieTweetings, Tripadvisor, Yahoo! and Yelp. Table~\ref{tab:cf_data} presents all domains and datasets used and a summary of their characteristics.

\begin{table*}[!ht]	
\centering
\small 
\caption{Summary of the datasets used in the experiments. Values within square brackets indicate lower and upper bounds in a specific characteristic. Notice that $k$ and $M$ stand for thousands and millions, respectively.} \label{tab:cf_data}
\begin{tabular}{llcccl}
\hline
	Domain & Dataset(s) & \#Users & \#Items & \#Ratings & Ref.\\ \hline
	Amazon & \makecell[cl]{App, Auto, Baby, Beauty, CD, Clothes, Food, Game, Garden, Health, Home, \\Instrument, Kindle, Movie, Music, Office, Pet, Phone, Sport, Tool, Toy, Video } & [7k - 311k] & [2k - 267k] & [11k - 574k] & \cite{McAuley2013} \\ \hline
	Bookcrossing & Bookcrossing & 8k & 29k & 40k & \cite{Ziegler2005}\\ \hline
	Flixter  & Flixter  & 15k & 22k & 813k & \cite{Zafarani+Liu:2009} \\ \hline
	Jester  & Jester1, Jester2, Jester3  & [2.3k - 2.5k] & [96 - 100] & [61k - 182k] & \cite{Goldberg2001} \\ \hline
	Movielens & \makecell[cl]{100k, 1m, 10m, 20m, latest}  & [94 - 23k] & [1k - 17k] & [10k - 2M] & \cite{GroupLens2016} \\ \hline
	MovieTweetings & RecSys2014, latest  & [2.5k - 3.7k] & [4.8k - 7.4k] & [21k - 39k] & \cite{Dooms13crowdrec} \\ \hline
	Tripadvisor  & Tripadvisor  & 78k & 11k & 151k & \cite{Wang2011} \\ \hline
	Yahoo! & Movies, Music  & [613 - 764] & [4k - 4.6k] & [22k - 31k] & \cite{Yahoo} \\ \hline
	Yelp  & Yelp  & 55k & 46k & 212k & \cite{Yelp2016}\\ \hline
\end{tabular}
\end{table*}

The CF algorithms used in this work are variations of MF methods: BPRMF~\cite{Rendle2009}, which performs a pairwise classification task, optimizing AUC using Stochastic Gradient Descent (SGD); WBPRMF~\cite{Rendle2009}, which is a variation of BPRMF that includes a sampling mechanism that promotes low scored items; SMRMF~\cite{Weimer2008}, which is another variation of BPRMF, but it replaces the optimization formula in SGD by a soft margin ranking loss inspired by SVM classifiers; WRMF~\cite{Hu2008a} which uses ALS (Alternating Least Squares) instead of SGD and introduces user/item bias to regularize the process; and lastly the baseline algorithm MostPopular which ranks items by how often they have been seen in the past. Since these algorithm tackle a Top-N recommendation problem, all algorithms are evaluated using NDCG (to assess ranking accuracy) and AUC (to evaluate classification accuracy) using 10-fold cross-validation. No parameter optimization was done to prevent bias towards any algorithm.

\vspace{4cm}

\subsection{Metalevel}

CF4CF uses only algorithm performance as input data. While the results obtained from the baselevel are used as training data, the prediction stage requires to calculate subsampling landmarkers. To do so, all datasets are random sampled for 10\% of all instances. Then, these samples are submitted to the same baselevel evaluation procedure to obtain performance estimations for all algorithms in all evaluation measures. In the case of MtL, each dataset is simply described by the state of the art metafeatures~\cite{Cunha2016} presented in Section~\ref{sec:meta_cf}. The algorithm performance is used to create rankings of algorithms to be used as targets for this predictive procedure. This means MtL is addressed using Label Ranking (LR)~\cite{Hullermeier2008,Vembu2010}. Recall that CF4CF is designed to use any CF algorithm. However, in order to provide the fairest comparison possible between MtL and CF, this work uses two algorithms with the same bias: user-based CF~\cite{Sarwar2000} and kNN for LR~\cite{Soares2015}, both based on Nearest Neighbours. These algorithms are referred to as KNN-CF and KNN-LR.

The evaluation in algorithm selection is comprised of two tasks: meta-accuracy and impact on the baselevel performance. While the first aims to assess how similar are the predicted and real rankings of algorithms, the second investigates how the algorithms recommended by the metamodels actually perform on average for all datasets. To assess the meta-accuracy, this work uses the ranking accuracy measure Kendall's Tau using leave-one-out cross-validation. To assess the impact on the baselevel, the analysis calculates the average performance for different thresholds $t$. These thresholds refer to the number of algorithms from the predicted ranking which are considered for analysis. Hence, if $t=1$, only the first recommended algorithm is used. On the other hand, if $t=2$, then both the first and second algorithms are used. In this situation, the performance is the best of both recommended algorithms. 

\section{Results}\label{sec:results}

\subsection{Rating Matrix Sparsity}\label{sub:sparsity_analysis}

\begin{figure}[!ht]
  \centering
    \includegraphics[width=.9\linewidth,trim={0 0.75cm 0 0},clip]{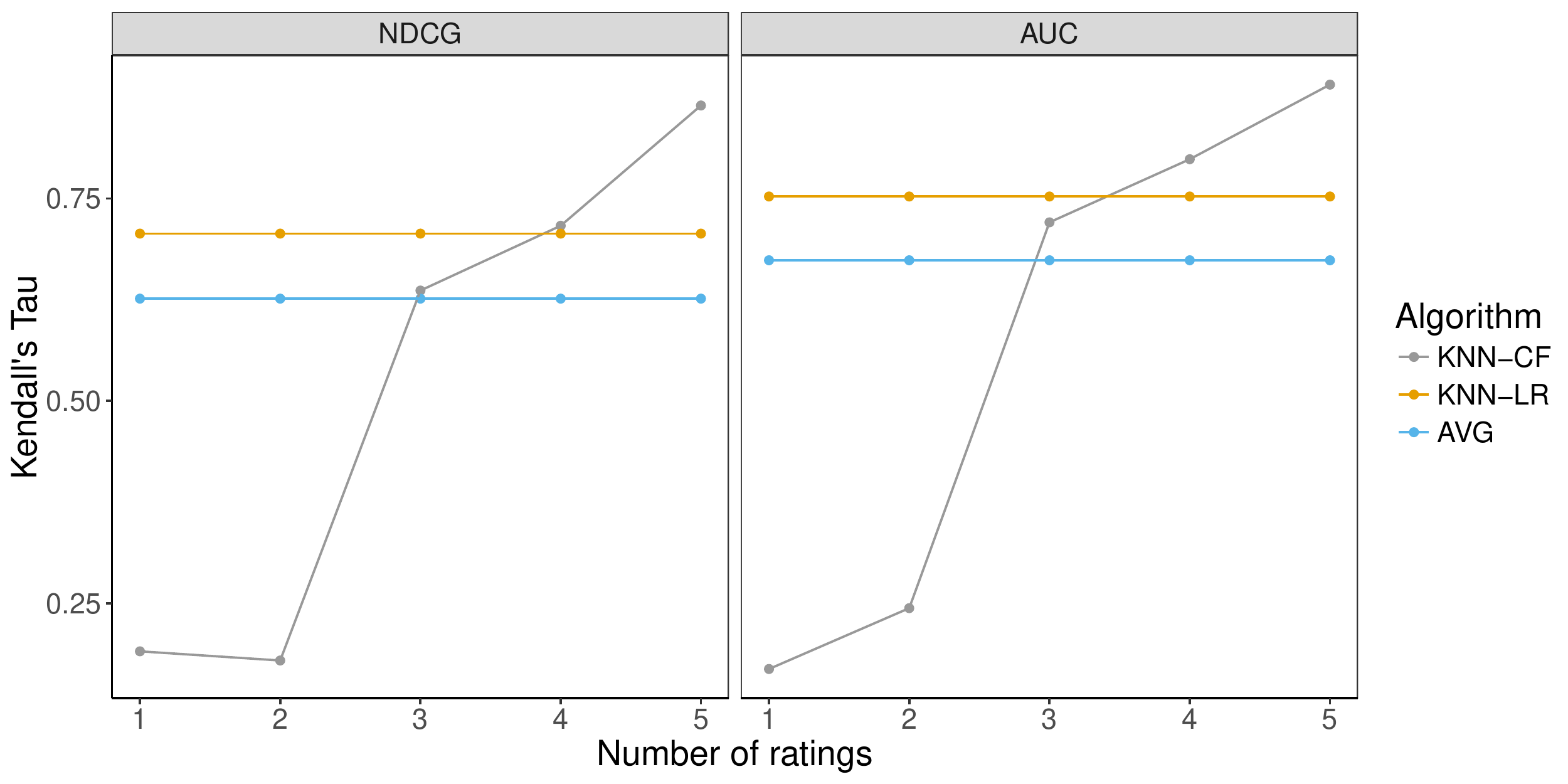}
      \caption{Ranking accuracy for different $N_{ratings}$.}
      \label{fig:sparsity}
\end{figure}

The first analysis aims at understanding the effect of variable $N_{ratings}$. To do so, different matrices were created by sampling the complete matrix and then CF4CF models were trained upon them. The results in terms of Kendall's Tau are presented in Figure~\ref{fig:sparsity}.

The results show CF4CF is equal or better than the baseline and MtL for $N_{ratings} = 3$ and $N_{ratings} = 4$, respectively. This shows CF4CF is able to provide good recommendations using only 4 ratings per baselevel dataset. However, the results also show that CF4CF is only better than MtL for $N_{ratings} = 5$ , meaning the full rating matrix is the only to consistently beat MtL. To obtain optimal results and provide fair comparison against MtL, we use a complete rating matrix in the remaining experiments.

\subsection{Meta-accuracy}\label{sub:new_user_ratings}

\begin{figure}[!ht]
  \centering
    \includegraphics[width=.9\linewidth,trim={0 0.7cm 0 0},clip]{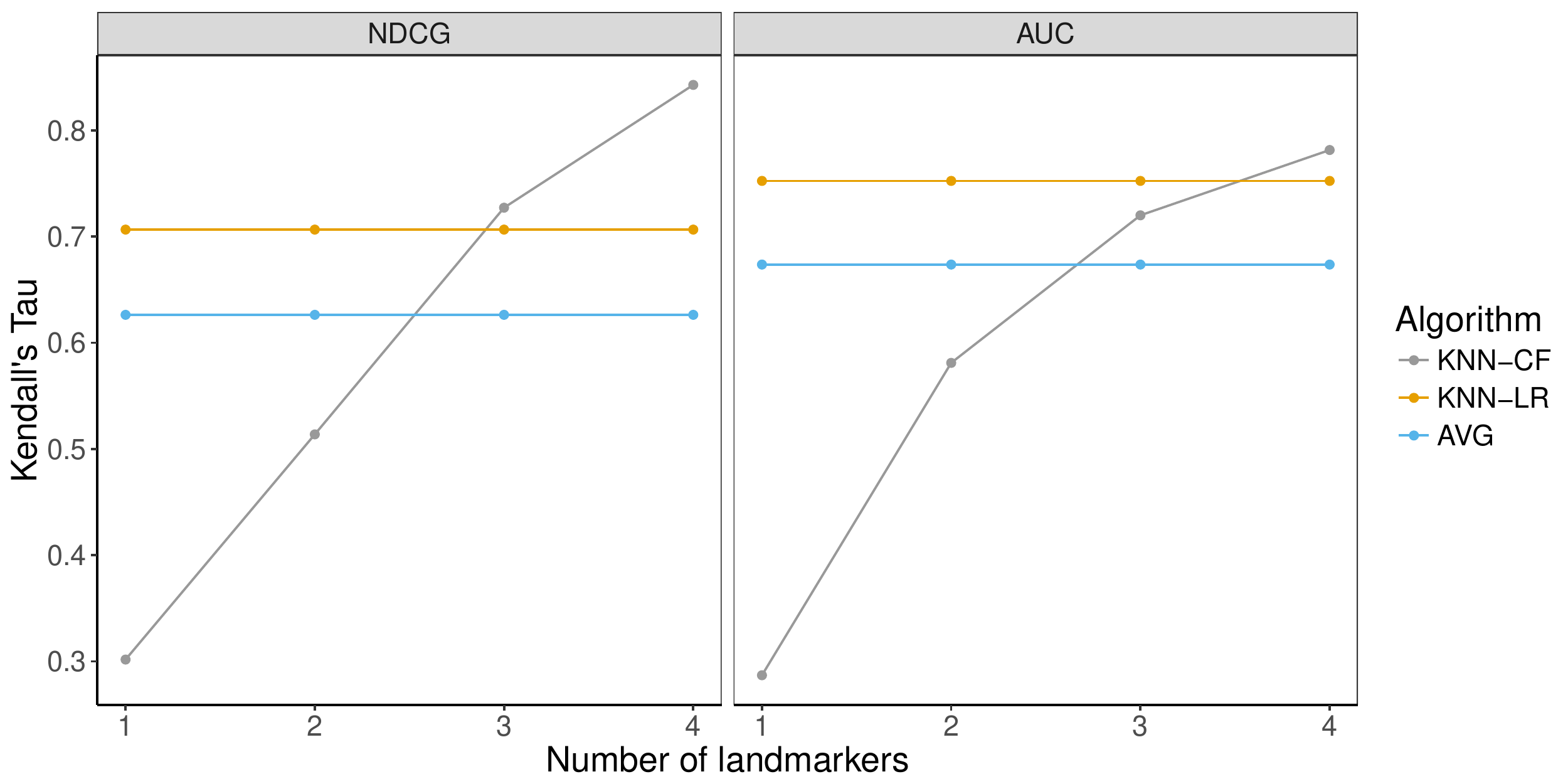}
      \caption{Ranking accuracy for different $N_{SL}$.}
      \label{fig:ratings}
\end{figure}

This analysis assesses the effect that the number of sampled landmarkers ($N_{SL}$) has in the overall performance of CF4CF. The Kendall's Tau results are presented in Figure~\ref{fig:ratings}. 

The results CF4CF is better than the baseline for $N_{SL}=3$ for both NDCG and AUC metatargets, but it only reaches comparable performance with regards to MtL for $N_{SL}=3$ in NDCG and $N_{SL}=4$ in AUC. Furthermore, CF4CF can outperform MtL but only for NDCG for $N_{SL}=4$. This means CF4CF is a suitable alternative to MtL, which in fact can perform better when 4 subsampling landmarkers are used to feed the CF metamodel.

\subsection{Impact on the baselevel performance}\label{sub:base_impact_cf}

\begin{figure}[!ht]
  \centering
    \includegraphics[width=.9\linewidth,trim={0 0.7cm 0 0},clip]{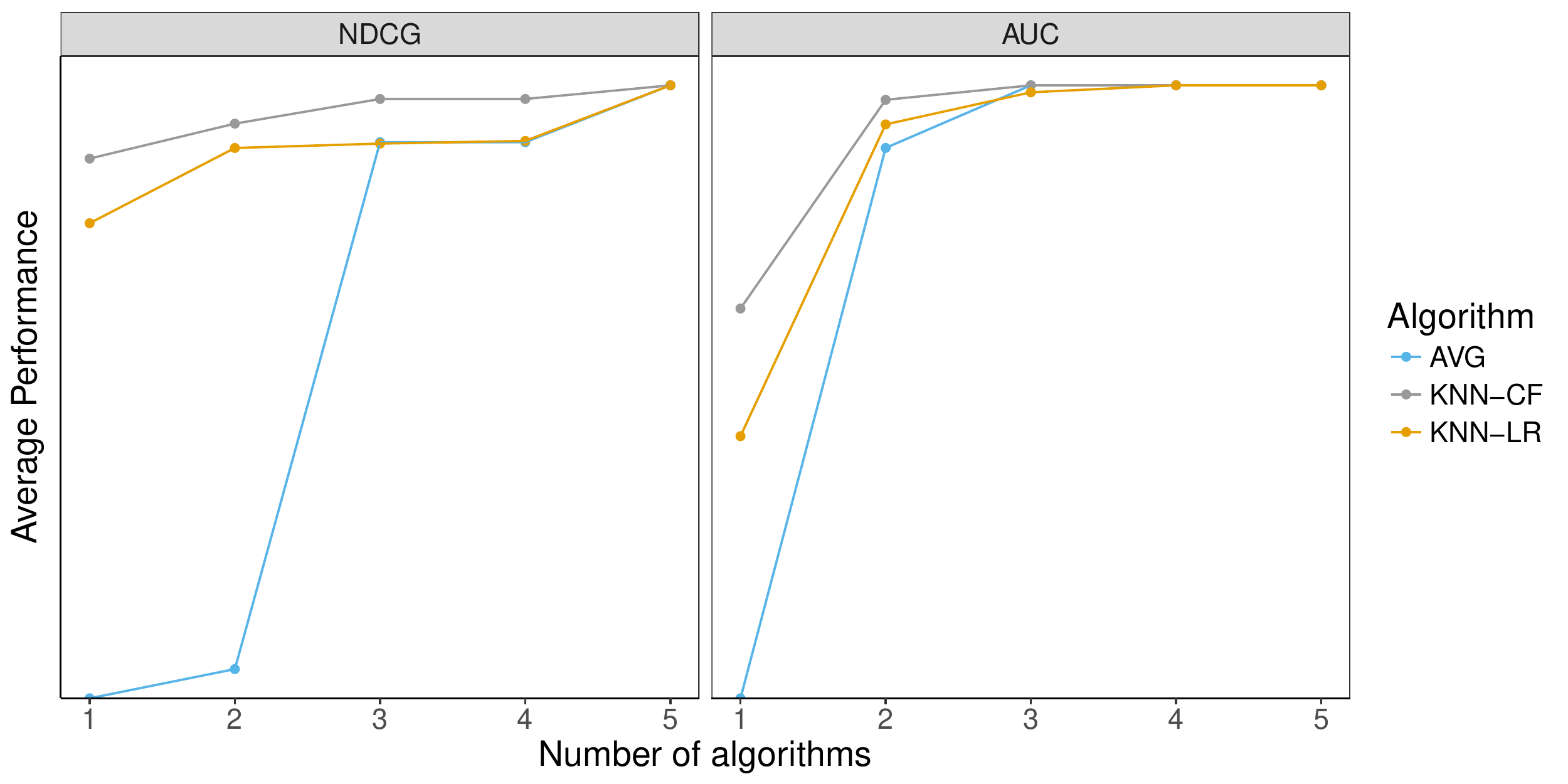}
      \caption{Impact on the baselevel performance.}
      \label{fig:base_performance}
\end{figure}

The results the impact on the baselevel performance are presented in Figure~\ref{fig:base_performance}. Notice the results presented refer to $N_{SL}=4$. 

The experimental results show CF4CF outperforms both the baseline and MtL for $t \in \{1,2,3,4\}$ and $t \in \{1,2\}$ for the the NDCG and AUC metatargets, respectively. These results show CF4CF makes better predictions than the competing approaches for the first thresholds in each problem, i.e. CF4CF is more accurate than MtL for the top positions in the predicted rankings of algorithms. 

\section{Conclusions}\label{sec:conclusions}

This work introduced a novel algorithm selection approach - CF4CF - which takes advantage of a Collaborative Filtering to recommend rankings of Collaborative Filtering algorithms. The procedure uses the algorithm performance as rating information to train the metamodel and uses subsampling landmarkers converted into ratings in the prediction stage. The proposed approach is the first known solution of its kind. According to the experimental results, CF4CF is a good alternative to MtL, and even better in some cases. CF4CF is able to perform equally to MtL using less data from algorithm performance in the rating matrix; it can out perform MtL when using 4 subsampling landmarkers in conjunction with a CF model; and it is able to have higher impact in the rankings of algorithms recommended in the top positions. All these observations allow to conclude CF4CF is better at predicting rankings of CF algorithms, (2) the CF algorithm it recommends has higher impact on the baselevel performance and (3) subsampling landmarkers are a suitable solution to provide initial ratings. Future work directions include: to improve CF4CF performance by testing different ways to leverage data for training and testing, further extend the experimental setup to other recommendation areas and algorithms and to leverage both metafeatures and ratings in a hybrid solution for CF algorithm selection.

\subsubsection*{Acknowledgments}

This work is financed by the Portuguese funding institution FCT - Fundação para a Ciência e a Tecnologia through the PhD grant SFRH/BD/117531/2016.

\clearpage 

\bibliographystyle{ACM-Reference-Format}
\bibliography{myrefs}


\begin{thebibliography}{26}


\ifx \showCODEN    \undefined \def \showCODEN     #1{\unskip}     \fi
\ifx \showDOI      \undefined \def \showDOI       #1{#1}\fi
\ifx \showISBNx    \undefined \def \showISBNx     #1{\unskip}     \fi
\ifx \showISBNxiii \undefined \def \showISBNxiii  #1{\unskip}     \fi
\ifx \showISSN     \undefined \def \showISSN      #1{\unskip}     \fi
\ifx \showLCCN     \undefined \def \showLCCN      #1{\unskip}     \fi
\ifx \shownote     \undefined \def \shownote      #1{#1}          \fi
\ifx \showarticletitle \undefined \def \showarticletitle #1{#1}   \fi
\ifx \showURL      \undefined \def \showURL       {\relax}        \fi
\providecommand\bibfield[2]{#2}
\providecommand\bibinfo[2]{#2}
\providecommand\natexlab[1]{#1}
\providecommand\showeprint[2][]{arXiv:#2}

\bibitem[\protect\citeauthoryear{Adomavicius and Zhang}{Adomavicius and
  Zhang}{2012}]%
        {Adomavicius2012}
\bibfield{author}{\bibinfo{person}{Gediminas Adomavicius} {and}
  \bibinfo{person}{Jingjing Zhang}.} \bibinfo{year}{2012}\natexlab{}.
\newblock \showarticletitle{{Impact of data characteristics on recommender
  systems performance}}.
\newblock \bibinfo{journal}{\emph{ACM Management Information Systems}}
  \bibinfo{volume}{3}, \bibinfo{number}{1} (\bibinfo{year}{2012}),
  \bibinfo{pages}{1--17}.
\newblock


\bibitem[\protect\citeauthoryear{Cunha, Soares, and Carvalho}{Cunha
  et~al\mbox{.}}{2017a}]%
        {Cunha:2017:MCF:3109859.3109899}
\bibfield{author}{\bibinfo{person}{Tiago Cunha}, \bibinfo{person}{Carlos
  Soares}, {and} \bibinfo{person}{Andr{\'{e}}~C.P.L.F. Carvalho}.}
  \bibinfo{year}{2017}\natexlab{a}.
\newblock \showarticletitle{{Metalearning for Context-aware Filtering:
  Selection of Tensor Factorization Algorithms}}. In
  \bibinfo{booktitle}{\emph{Proceedings of the Eleventh ACM Conference on
  Recommender Systems}} \emph{(\bibinfo{series}{RecSys '17})}.
  \bibinfo{publisher}{ACM}, \bibinfo{address}{New York, NY, USA},
  \bibinfo{pages}{14--22}.
\newblock
\showISBNx{978-1-4503-4652-8}
\urldef\tempurl%
\url{https://doi.org/10.1145/3109859.3109899}
\showDOI{\tempurl}


\bibitem[\protect\citeauthoryear{Cunha, Soares, and de~Carvalho}{Cunha
  et~al\mbox{.}}{2016}]%
        {Cunha2016}
\bibfield{author}{\bibinfo{person}{Tiago Cunha}, \bibinfo{person}{Carlos
  Soares}, {and} \bibinfo{person}{Andr{\'{e}} de Carvalho}.}
  \bibinfo{year}{2016}\natexlab{}.
\newblock \showarticletitle{{Selecting Collaborative Filtering algorithms using
  Metalearning}}. In \bibinfo{booktitle}{\emph{ECML-PKDD}}.
  \bibinfo{pages}{393--409}.
\newblock


\bibitem[\protect\citeauthoryear{Cunha, Soares, and de~Carvalho}{Cunha
  et~al\mbox{.}}{2017b}]%
        {Cunha2017}
\bibfield{author}{\bibinfo{person}{Tiago Cunha}, \bibinfo{person}{Carlos
  Soares}, {and} \bibinfo{person}{Andre de Carvalho}.}
  \bibinfo{year}{2017}\natexlab{b}.
\newblock \showarticletitle{{Recommending Collaborative Filtering algorithms
  using subsampling landmarkers}}. In \bibinfo{booktitle}{\emph{Discovery
  Science}}. \bibinfo{pages}{189--203}.
\newblock


\bibitem[\protect\citeauthoryear{Cunha, Soares, and de~Carvalho}{Cunha
  et~al\mbox{.}}{2018}]%
        {Cunha2018128}
\bibfield{author}{\bibinfo{person}{Tiago Cunha}, \bibinfo{person}{Carlos
  Soares}, {and} \bibinfo{person}{Andr{\'{e}}~C.P.L.F. de Carvalho}.}
  \bibinfo{year}{2018}\natexlab{}.
\newblock \showarticletitle{{Metalearning and Recommender Systems: A literature
  review and empirical study on the algorithm selection problem for
  Collaborative Filtering}}.
\newblock \bibinfo{journal}{\emph{Information Sciences}}  \bibinfo{volume}{423}
  (\bibinfo{year}{2018}), \bibinfo{pages}{128--144}.
\newblock
\showISSN{0020-0255}


\bibitem[\protect\citeauthoryear{Dooms, {De Pessemier}, and Martens}{Dooms
  et~al\mbox{.}}{2013}]%
        {Dooms13crowdrec}
\bibfield{author}{\bibinfo{person}{Simon Dooms}, \bibinfo{person}{Toon {De
  Pessemier}}, {and} \bibinfo{person}{Luc Martens}.}
  \bibinfo{year}{2013}\natexlab{}.
\newblock \showarticletitle{{MovieTweetings: a Movie Rating Dataset Collected
  From Twitter}}. In \bibinfo{booktitle}{\emph{CrowdRec at RecSys 2013}}.
\newblock


\bibitem[\protect\citeauthoryear{Ekstrand and Riedl}{Ekstrand and
  Riedl}{2012}]%
        {Ekstrand2012}
\bibfield{author}{\bibinfo{person}{Michael Ekstrand} {and}
  \bibinfo{person}{John Riedl}.} \bibinfo{year}{2012}\natexlab{}.
\newblock \showarticletitle{{When Recommenders Fail: Predicting Recommender
  Failure for Algorithm Selection and Combination}}.
\newblock \bibinfo{journal}{\emph{ACM RecSys}} (\bibinfo{year}{2012}),
  \bibinfo{pages}{233--236}.
\newblock
\showISBNx{9781450312707}


\bibitem[\protect\citeauthoryear{Goldberg, Roeder, Gupta, and Perkins}{Goldberg
  et~al\mbox{.}}{2001}]%
        {Goldberg2001}
\bibfield{author}{\bibinfo{person}{Ken Goldberg}, \bibinfo{person}{Theresa
  Roeder}, \bibinfo{person}{Dhruv Gupta}, {and} \bibinfo{person}{Chris
  Perkins}.} \bibinfo{year}{2001}\natexlab{}.
\newblock \showarticletitle{{Eigentaste: A Constant Time Collaborative
  Filtering Algorithm}}.
\newblock \bibinfo{journal}{\emph{Information Retrieval}} \bibinfo{volume}{4},
  \bibinfo{number}{2} (\bibinfo{year}{2001}), \bibinfo{pages}{133--151}.
\newblock


\bibitem[\protect\citeauthoryear{Griffith, O'Riordan, and Sorensen}{Griffith
  et~al\mbox{.}}{2012}]%
        {Griffith2012}
\bibfield{author}{\bibinfo{person}{Josephine Griffith}, \bibinfo{person}{Colm
  O'Riordan}, {and} \bibinfo{person}{Humphrey Sorensen}.}
  \bibinfo{year}{2012}\natexlab{}.
\newblock \showarticletitle{{Investigations into user rating information and
  accuracy in collaborative filtering}}. In \bibinfo{booktitle}{\emph{ACM
  SAC}}. \bibinfo{pages}{937--942}.
\newblock


\bibitem[\protect\citeauthoryear{GroupLens}{GroupLens}{2016}]%
        {GroupLens2016}
\bibfield{author}{\bibinfo{person}{GroupLens}.}
  \bibinfo{year}{2016}\natexlab{}.
\newblock \bibinfo{title}{{MovieLens datasets}}.
\newblock   (\bibinfo{year}{2016}).
\newblock
\urldef\tempurl%
\url{http://grouplens.org/datasets/movielens/}
\showURL{%
\tempurl}


\bibitem[\protect\citeauthoryear{Hu, Koren, and Volinsky}{Hu
  et~al\mbox{.}}{2008}]%
        {Hu2008a}
\bibfield{author}{\bibinfo{person}{Yifan Hu}, \bibinfo{person}{Yehuda Koren},
  {and} \bibinfo{person}{Chris Volinsky}.} \bibinfo{year}{2008}\natexlab{}.
\newblock \showarticletitle{{Collaborative Filtering for Implicit Feedback
  Datasets}}. In \bibinfo{booktitle}{\emph{IEEE International Conference on
  Data Mining}}. \bibinfo{pages}{263 -- 272}.
\newblock


\bibitem[\protect\citeauthoryear{H{\"{u}}llermeier, F{\"{u}}rnkranz, Cheng, and
  Brinker}{H{\"{u}}llermeier et~al\mbox{.}}{2008}]%
        {Hullermeier2008}
\bibfield{author}{\bibinfo{person}{Eyke H{\"{u}}llermeier},
  \bibinfo{person}{Johannes F{\"{u}}rnkranz}, \bibinfo{person}{Weiwei Cheng},
  {and} \bibinfo{person}{Klaus Brinker}.} \bibinfo{year}{2008}\natexlab{}.
\newblock \showarticletitle{{Label ranking by learning pairwise preferences}}.
\newblock \bibinfo{journal}{\emph{Artificial Intelligence}}
  \bibinfo{volume}{172}, \bibinfo{number}{16-17} (\bibinfo{year}{2008}),
  \bibinfo{pages}{1897--1916}.
\newblock


\bibitem[\protect\citeauthoryear{Matuszyk and Spiliopoulou}{Matuszyk and
  Spiliopoulou}{2014}]%
        {Matuszyk2014}
\bibfield{author}{\bibinfo{person}{Pawel Matuszyk} {and} \bibinfo{person}{Myra
  Spiliopoulou}.} \bibinfo{year}{2014}\natexlab{}.
\newblock \showarticletitle{{Predicting the Performance of Collaborative
  Filtering Algorithms}}. In \bibinfo{booktitle}{\emph{Web Intelligence, Mining
  and Semantics}}. \bibinfo{pages}{38:1--38:6}.
\newblock


\bibitem[\protect\citeauthoryear{McAuley and Leskovec}{McAuley and
  Leskovec}{2013}]%
        {McAuley2013}
\bibfield{author}{\bibinfo{person}{Julian McAuley} {and} \bibinfo{person}{Jure
  Leskovec}.} \bibinfo{year}{2013}\natexlab{}.
\newblock \showarticletitle{{Hidden Factors and Hidden Topics: Understanding
  Rating Dimensions with Review Text}}. In \bibinfo{booktitle}{\emph{ACM
  Conference on Recommender Systems}}. \bibinfo{pages}{165--172}.
\newblock


\bibitem[\protect\citeauthoryear{Pinto, Soares, and Mendes-Moreira}{Pinto
  et~al\mbox{.}}{2016}]%
        {Pinto2016}
\bibfield{author}{\bibinfo{person}{F{\'{a}}bio Pinto}, \bibinfo{person}{Carlos
  Soares}, {and} \bibinfo{person}{Jo{\~{a}}o Mendes-Moreira}.}
  \bibinfo{year}{2016}\natexlab{}.
\newblock \showarticletitle{{Towards automatic generation of Metafeatures}}. In
  \bibinfo{booktitle}{\emph{PAKDD}}. \bibinfo{pages}{215--226}.
\newblock


\bibitem[\protect\citeauthoryear{Rendle, Freudenthaler, Gantner, and
  Schmidt-thieme}{Rendle et~al\mbox{.}}{2009}]%
        {Rendle2009}
\bibfield{author}{\bibinfo{person}{Steffen Rendle}, \bibinfo{person}{Christoph
  Freudenthaler}, \bibinfo{person}{Zeno Gantner}, {and} \bibinfo{person}{Lars
  Schmidt-thieme}.} \bibinfo{year}{2009}\natexlab{}.
\newblock \showarticletitle{{BPR: Bayesian Personalized Ranking from Implicit
  Feedback}}. In \bibinfo{booktitle}{\emph{Proceedings of the Twenty-Fifth
  Conference on Uncertainty in Artificial Intelligence}}.
  \bibinfo{pages}{452--461}.
\newblock


\bibitem[\protect\citeauthoryear{Sarwar, Karypis, Konstan, and Riedl}{Sarwar
  et~al\mbox{.}}{2000}]%
        {Sarwar2000}
\bibfield{author}{\bibinfo{person}{Badrul Sarwar}, \bibinfo{person}{George
  Karypis}, \bibinfo{person}{Joseph Konstan}, {and} \bibinfo{person}{John
  Riedl}.} \bibinfo{year}{2000}\natexlab{}.
\newblock \showarticletitle{{Analysis of Recommendation Algorithms for
  E-Commerce}}. In \bibinfo{booktitle}{\emph{ACM Electronic Commerce}}.
  \bibinfo{pages}{158--167}.
\newblock
\showISBNx{1581132727}


\bibitem[\protect\citeauthoryear{Shi, Larson, and Hanjalic}{Shi
  et~al\mbox{.}}{2014}]%
        {Shi2014}
\bibfield{author}{\bibinfo{person}{Yue Shi}, \bibinfo{person}{Martha Larson},
  {and} \bibinfo{person}{Alan Hanjalic}.} \bibinfo{year}{2014}\natexlab{}.
\newblock \showarticletitle{{Collaborative Filtering beyond the User-Item
  Matrix}}.
\newblock \bibinfo{journal}{\emph{Comput. Surveys}} \bibinfo{volume}{47},
  \bibinfo{number}{1} (\bibinfo{year}{2014}), \bibinfo{pages}{1--45}.
\newblock


\bibitem[\protect\citeauthoryear{Soares}{Soares}{2015}]%
        {Soares2015}
\bibfield{author}{\bibinfo{person}{Carlos Soares}.}
  \bibinfo{year}{2015}\natexlab{}.
\newblock \bibinfo{title}{{labelrank: Predicting Rankings of Labels}}.
\newblock   (\bibinfo{year}{2015}).
\newblock
\urldef\tempurl%
\url{https://cran.r-project.org/package=labelrank}
\showURL{%
\tempurl}


\bibitem[\protect\citeauthoryear{Vembu and G{\"{a}}rtner}{Vembu and
  G{\"{a}}rtner}{2010}]%
        {Vembu2010}
\bibfield{author}{\bibinfo{person}{Shankar Vembu} {and} \bibinfo{person}{Thomas
  G{\"{a}}rtner}.} \bibinfo{year}{2010}\natexlab{}.
\newblock \showarticletitle{{Label ranking algorithms: A survey}}.
\newblock In \bibinfo{booktitle}{\emph{Preference Learning}}.
  \bibinfo{pages}{45--64}.
\newblock


\bibitem[\protect\citeauthoryear{Wang, Lu, and Zhai}{Wang
  et~al\mbox{.}}{2011}]%
        {Wang2011}
\bibfield{author}{\bibinfo{person}{Hongning Wang}, \bibinfo{person}{Yue Lu},
  {and} \bibinfo{person}{ChengXiang Zhai}.} \bibinfo{year}{2011}\natexlab{}.
\newblock \showarticletitle{{Latent Aspect Rating Analysis Without Aspect
  Keyword Supervision}}. In \bibinfo{booktitle}{\emph{ACM SIGKDD}}.
  \bibinfo{pages}{618--626}.
\newblock


\bibitem[\protect\citeauthoryear{Weimer, Karatzoglou, and Smola}{Weimer
  et~al\mbox{.}}{2008}]%
        {Weimer2008}
\bibfield{author}{\bibinfo{person}{Markus Weimer}, \bibinfo{person}{Alexandros
  Karatzoglou}, {and} \bibinfo{person}{Alex Smola}.}
  \bibinfo{year}{2008}\natexlab{}.
\newblock \showarticletitle{{Improving Maximum Margin Matrix Factorization}}.
\newblock \bibinfo{journal}{\emph{Machine Learning}} \bibinfo{volume}{72},
  \bibinfo{number}{3} (\bibinfo{year}{2008}), \bibinfo{pages}{263--276}.
\newblock


\bibitem[\protect\citeauthoryear{Yahoo!}{Yahoo!}{2016}]%
        {Yahoo}
\bibfield{author}{\bibinfo{person}{Yahoo!}} \bibinfo{year}{2016}\natexlab{}.
\newblock \bibinfo{title}{{Webscope datasets}}.
\newblock   (\bibinfo{year}{2016}).
\newblock
\urldef\tempurl%
\url{https://webscope.sandbox.yahoo.com/}
\showURL{%
\tempurl}


\bibitem[\protect\citeauthoryear{Yelp}{Yelp}{2016}]%
        {Yelp2016}
\bibfield{author}{\bibinfo{person}{Yelp}.} \bibinfo{year}{2016}\natexlab{}.
\newblock \bibinfo{title}{{Yelp Dataset Challenge}}.
\newblock   (\bibinfo{year}{2016}).
\newblock
\urldef\tempurl%
\url{https://www.yelp.com/dataset\_challenge}
\showURL{%
\tempurl}


\bibitem[\protect\citeauthoryear{Zafarani and Liu}{Zafarani and Liu}{2009}]%
        {Zafarani+Liu:2009}
\bibfield{author}{\bibinfo{person}{R. Zafarani} {and} \bibinfo{person}{H.
  Liu}.} \bibinfo{year}{2009}\natexlab{}.
\newblock \bibinfo{title}{Social Computing Data Repository at {ASU}}.
\newblock   (\bibinfo{year}{2009}).
\newblock
\urldef\tempurl%
\url{http://socialcomputing.asu.edu}
\showURL{%
\tempurl}


\bibitem[\protect\citeauthoryear{Ziegler, McNee, Konstan, and Lausen}{Ziegler
  et~al\mbox{.}}{2005}]%
        {Ziegler2005}
\bibfield{author}{\bibinfo{person}{Cai-Nicolas Ziegler},
  \bibinfo{person}{Sean~M McNee}, \bibinfo{person}{Joseph~A Konstan}, {and}
  \bibinfo{person}{Georg Lausen}.} \bibinfo{year}{2005}\natexlab{}.
\newblock \showarticletitle{{Improving Recommendation Lists Through Topic
  Diversification}}. In \bibinfo{booktitle}{\emph{Proceedings of the 14th
  International Conference on World Wide Web}}. \bibinfo{pages}{22--32}.
\newblock


\end{thebibliography}

\end{document}